\documentclass[reprint, amsmath,amssymb, aps, superscriptaddress, floatfix]{revtex4-1}

\usepackage[utf8]{inputenc}
\usepackage{graphicx}
\usepackage{dcolumn}
\usepackage{bm}
\usepackage{braket}
\usepackage{color}
\usepackage{units}
\usepackage{dsfont}
\usepackage{float}
\usepackage{hyperref}
\usepackage{xcolor}
\usepackage{comment}

\begin{document}

\title{Ultrafast switching of telecom photon-number states}

\author{Kate L. Fenwick}
\affiliation{National Research Council of Canada, 100 Sussex Drive, Ottawa, Ontario K1A 0R6, Canada}
\author{Fr\'ed\'eric Bouchard}
\email{frederic.bouchard@nrc-cnrc.gc.ca}
\affiliation{National Research Council of Canada, 100 Sussex Drive, Ottawa, Ontario K1A 0R6, Canada}
\author{Alicia Sit}
\affiliation{National Research Council of Canada, 100 Sussex Drive, Ottawa, Ontario K1A 0R6, Canada}
\author{Timothy Lee}
\affiliation{Department of Physics, University of Ottawa, Advanced Research Complex, 25 Templeton Street, Ottawa ON Canada, K1N 6N5}
\author{Andrew H. Proppe}
\affiliation{National Research Council of Canada, 100 Sussex Drive, Ottawa, Ontario K1A 0R6, Canada}
\affiliation{Department of Physics, University of Ottawa, Advanced Research Complex, 25 Templeton Street, Ottawa ON Canada, K1N 6N5}
\author{Guillaume Thekkadath}
\affiliation{National Research Council of Canada, 100 Sussex Drive, Ottawa, Ontario K1A 0R6, Canada}
\author{Duncan England}
\affiliation{National Research Council of Canada, 100 Sussex Drive, Ottawa, Ontario K1A 0R6, Canada}
\author{Philip J. Bustard}
\affiliation{National Research Council of Canada, 100 Sussex Drive, Ottawa, Ontario K1A 0R6, Canada}
\author{Jeff S. Lundeen}
\affiliation{Department of Physics, University of Ottawa, Advanced Research Complex, 25 Templeton Street, Ottawa ON Canada, K1N 6N5}
\author{Benjamin J. Sussman}
\affiliation{National Research Council of Canada, 100 Sussex Drive, Ottawa, Ontario K1A 0R6, Canada}
\affiliation{Department of Physics, University of Ottawa, Advanced Research Complex, 25 Templeton Street, Ottawa ON Canada, K1N 6N5}

\begin{abstract}
A crucial component of photonic quantum information processing platforms is the ability to modulate, route, convert, and switch quantum states of light noiselessly with low insertion loss. For instance, a high-speed, low-loss optical switch is crucial for scaling quantum photonic systems that rely on measurement-based feed-forward approaches. Such a device will also ideally be capable of operating on photon-number states, which can act as non-Gaussian resources. Here, we demonstrate ultrafast all-optical switching of heralded photon-number states, of up to 6 photons, using the optical Kerr effect in a single-mode fiber. A local birefringence is created by a high-intensity pump pulse at a center wavelength of 1030\,nm which overlaps temporally with the 1550\,nm photons in the fiber. A switching efficiency of $>$99\% is reached with a resolution of 2.3\,ps, {an insertion loss of $2.27\pm0.08$\,dB}, and a signal-to-noise ratio of 32,000.
\end{abstract}

\maketitle

The ability to modulate optical signals at high speeds has been a foundational technology in the telecommunications industry and a cornerstone of fundamental research in atomic, molecular, and optical physics. As we enter the next era of quantum photonic technologies, the design and implementation of fast optical modulators for single photons and quantum states of light will become increasingly crucial~\cite{prevedel2007high,mikova2012increasing,shomroni2014all,kaneda2019high,meyer2022scalable}. These new tools for quantum photonics will demand far more stringent specifications due to the significant impact that loss and noise can have on the performance of quantum applications. One key area of advancement lies in the development of single-photon modulators and switches capable of operating at speeds achievable only with fully optical control~\cite{hall2011ultrafast,hall2011all,kupchak2019terahertz}. In media with a third-order optical nonlinearity, all-optical switching leverages the high peak intensity of short pulses of light or long propagation length of optical fibers to access the nonlinear properties of a medium, through which a secondary pulse propagates. The nonlinearity is induced at the speed of the response of the electrons in the medium, which is much faster than current electronics. While gigahertz-frequency (nanosecond timescale) modulation is routine with electro-optic devices and fast electronics, the terahertz (picosecond timescale) regime remains accessible only through all-optical means at present. Pushing the capabilities of all-optical switching in the ultrafast regime resulting in switching times ranging from hundred femtoseconds to a few picoseconds, has resulted in many applications in ultrafast quantum information processing~\cite{bouchard2021achieving,bouchard2022quantum,fenwick2024photonic,bouchard2024programmable}.

Quantum technologies have reached a stage where scaling has become the primary challenge in their development and deployment. In photonic platforms, this challenge manifests in two key ways: first, the need to generate and detect quantum states with higher photon numbers or greater squeezing, and second, the necessity of mitigating loss, which remains a primary limitation in scaling quantum photonic systems~\cite{aghaee2025scaling,psiquantum2025manufacturable}. Ultrafast all-optical operation offers novel approaches to addressing the challenges of scaling quantum photonic systems to larger photon-number states. For example, operating at the picosecond timescale enables photon-number-resolving (PNR) capabilities in superconducting nanowire single-photon detectors (SNSPDs)~\cite{cahall2017multi,zhu2020resolving}. At these ultrafast timescales, the temporal characteristics of the detector’s output signal provide information about the incident photon number. In contrast, transition-edge sensors (TESs) offer intrinsic PNR capability independent of pulse duration, but their microsecond response time is a significant limitation~\cite{gerrits2016superconducting}. By integrating TESs with an ultrafast all-optical switch, their slow operation can be circumvented, enabling fast detection while preserving their superior photon-number resolution. Photon-number resolution is imperative for measuring the exemplar quantum state, the photon-number state. Here, we demonstrate that our all-optical switch is capable of operating on these important quantum states.

\begin{figure*}[t!]
	\centering
		\includegraphics[width=0.85\textwidth]{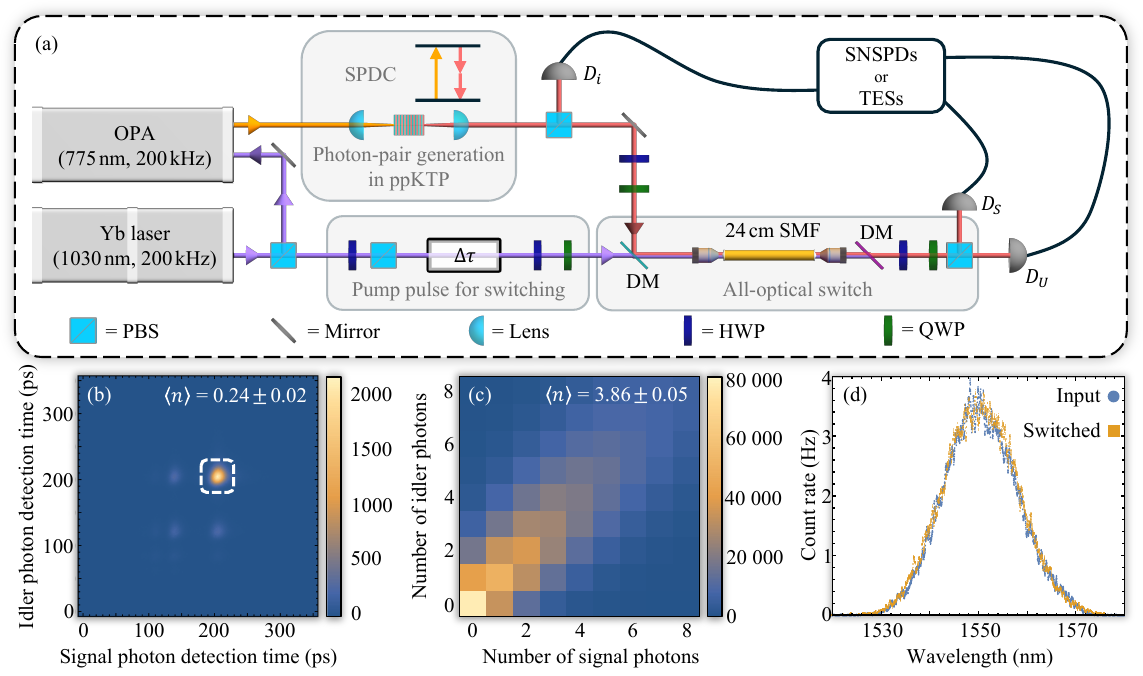}
	\caption{\textbf{Experimental setup.} A simplified experimental setup is shown in (a), where a pulsed Ytterbium (Yb) laser pumps an optical parametric amplifier (OPA) and provides the pump pulses necessary for ultrafast all-optical switching. Output pulses from the OPA undergo spontaneous parametric downconversion (SPDC) in a periodically poled potassium titanyl phosphase (ppKTP) crystal for photon-pair generation. The idler photons, measured at detector $D_i$, herald the detection of the switched (unswitched) signal photons at detector $D_S$ ($D_U$). Photons are detected by either superconducting nanowire single-photon detectors (SNSPDs) or transition-edge sensors (TESs). Extracting 2D timing histograms from the detection events measured by the SNSPDs, as shown in (b), allows us to filter the one-photon events (enclosed by a white dashed line corresponding to a 60\,ps correlation window) from events with more than one photon. The SPDC source is operated in the low mean photon number regime, $\braket{n}=0.24\pm0.02$, for the single-photon switching demonstrations made with the SNSPDs. On the other hand, the source is operated in the high mean photon number regime, $\braket{n}=3.86\pm0.05$, for the switching of heralded number states with up to 6 photons, as measured by the TESs. A joint photon-number intensity measurement is shown in (c) for the latter case. The signal photon spectrum remains unchanged by the switch, as shown in (d). DM: dichroic mirror; SMF: single-mode fiber; PBS: polarizing beamsplitter; HWP: half-wave plate; QWP: quarter-wave plate; $\Delta\tau$: pump delay line.}
	\label{fig:experimental}
\end{figure*}

Beyond detection, a key requirement for scaling quantum photonic systems is the implementation of measurement-based feed-forward operations. Many quantum protocols rely on real-time measurements to conditionally manipulate the remaining quantum state, demanding fast, low-loss switching and modulation. However, the processing of detector signals and subsequent feed-forward to an optical switch introduces latencies on the order of microseconds—timescales over which maintaining coherence in large photonic quantum states becomes increasingly challenging. Ultrafast operation provides a temporal encoding approach known as ultrafast time-bin encoding (UTBE)~\cite{bouchard2024programmable}. At the picosecond timescale, environmental fluctuations are effectively frozen, suppressing decoherence caused by mechanical instabilities and enabling high-fidelity quantum operations. This regime further allows long single-mode fibers to serve as low-loss quantum buffers~\cite{lee2024fiber}, preserving quantum states while feed-forward operations are executed. {Interestingly, the feed-forward operations necessary for quantum information processing need not be limited by electronics nor optoelectronics; recent demonstrations have shown that all-optical utrafast feed-forward operations~\cite{yamashima2025all} and single-photon detection~\cite{sendonaris2024ultrafast}, both based on optical parametric amplification, are possible.}The development of high-speed, low-loss all-optical switches and modulators is therefore essential for realizing scalable, real-time adaptive quantum photonic architectures.

Finally, operating in the ultrafast regime expands the quantum photonics toolkit by providing access to broad optical bandwidths. For instance, the large bandwidth of ultrashort pump pulses is leveraged in state-of-the-art heralded single-photon sources to achieve high spectral purity, while their high pulse energy enables stronger squeezing, supporting the generation of large heralded photon-number states~\cite{harder2016single}. Despite the advantages and promises of ultrafast quantum photonics, most demonstrations of all-optical switching have been concentrated in the visible regime (around 800 nm)~\cite{donohue2013coherent,kupchak2017time,bouchard2023measuring}, whereas the telecom band (1550 nm) remains relatively less explored. Evidence of low loss, ultrafast switching of telecom single photons has been reported~\cite{purakayastha2022ultrafast}, yet a high-efficiency implementation remains an open challenge, for which this work presents a solution. Advances in superconducting detection have enabled low-jitter, photon-number-resolving detectors optimized for telecom wavelengths, where optical fiber transmission losses are minimized. As quantum photonic technologies scale, integrating ultrafast all-optical components with these detectors will be important for realizing high-speed, low-loss quantum information processing systems.

In this work, we design and demonstrate a high-efficiency all-optical switch which operates on heralded photon-number states in the telecom band with picosecond resolution, low noise, and low loss. The switch relies on cross-phase modulation (XPM) via the optical Kerr effect in a short single-mode fiber (SMF)~\cite{england2021perspectives}. The Kerr effect, a third-order nonlinear process, induces an intensity-dependent change in the refractive index of the fiber due to the nonlinear polarization response of the material. For example, when a high-intensity pump pulse propagates through the fiber, it induces a local birefringence, altering the refractive index experienced by a co-propagating, weaker signal pulse. This XPM effect can be used to manipulate the polarization of the signal pulse, depending on the pump pulse intensity and polarization. Precise control over the pump pulse intensity, and the relative polarization and timing between the pump and signal pulses, allows for tunable all-optical switching. With access to ultrashort pulses (hundreds-of-femtoseconds in duration) and detectors with PNR capabilities, picosecond-resolution switching of photon-number states can be achieved.

Shown in Fig.~\ref{fig:experimental}(a), the experiment is driven by a commercial Ytterbium-doped ultrafast laser system (Carbide, LightConversion) delivering 180\,fs pulses at a center wavelength of 1030\,nm with a 200\,kHz repetition rate. The beam is split, with the majority directed to an optical parametric amplifier (Orpheus-HP, LightConversion) tuned to 775\,nm which pumps a 2\,mm type-II periodically poled potassium titanyl phosphate (ppKTP) crystal. Degenerate photon pairs at 1550\,nm are generated via spontaneous parametric down-conversion (SPDC), with the signal and idler photons separated by a polarizing beam splitter (PBS). The idler photons are detected (at $D_i$) to herald the signal photons, which are coupled into a 24\,cm SMF (1060XP, Thorlabs) and switched.

\begin{figure}[t!]
	\centering
		\includegraphics[width=0.45\textwidth]{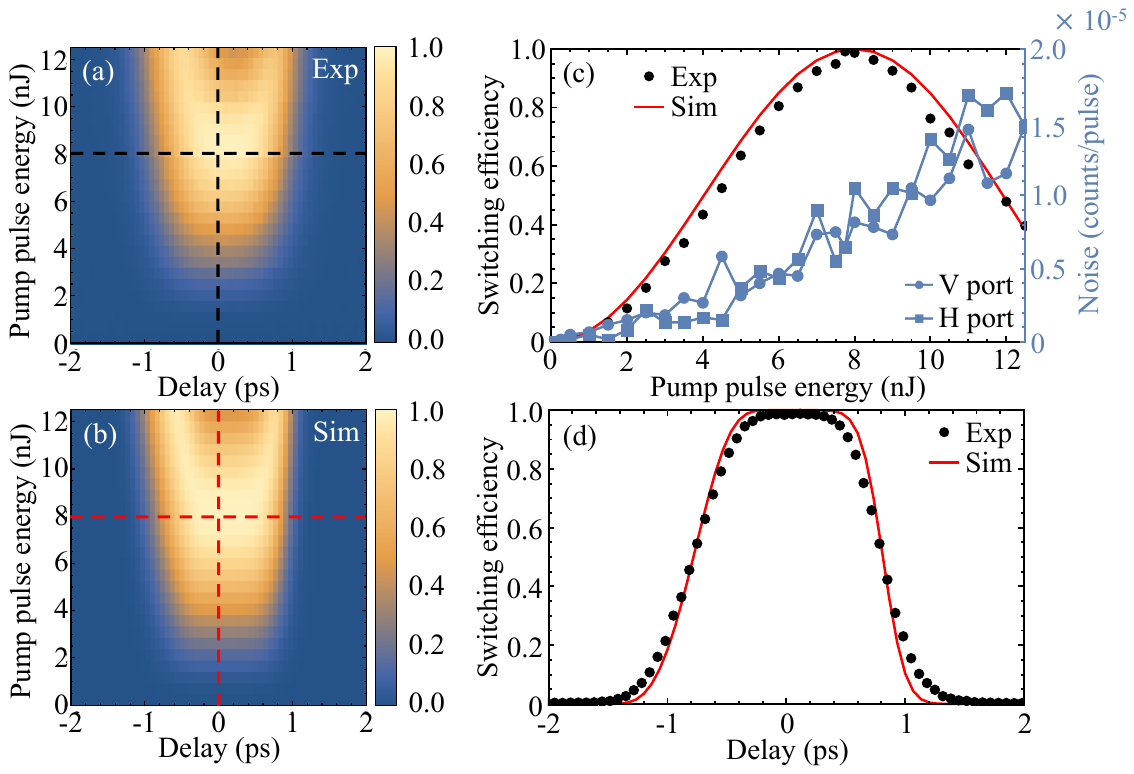}
	\caption{\textbf{Characterizing the switch.} The experimental and simulated (see Supplemental Material) switching efficiency, as a function of pump pulse energy and delay, are shown in (a) and (b), respectively. Slices from these 2D datasets are shown for (c) a constant pump pulse delay of $\Delta\tau=0$ and (d) a constant pump pulse energy of 8\,nJ, where the black circles correspond to experimental data and red lines correspond to simulation results. Also shown in (c) are the measured pump noise counts per pulse, indicated by the blue circles and squares for the switched (V) and unswitched (H) ports, respectively. {Note that error bars based on Poissonian statistics are smaller than the symbol size for all data points in (c) and (d).} Here, SNSPDs are used for experimental measurements.}
	\label{fig:SNSPDdata}
\end{figure}

Part of the remaining 1030\,nm beam is coupled into this 24\,cm SMF to pump the XPM process required for switching. Spatiotemporal overlap between the pump pulse and signal photons is achieved by recombining the beams on a dichroic mirror prior to the SMF, and adjusting their relative delay, $\Delta\tau$. The polarization of the pump pulse is precisely set to maximize the polarization rotation of the signal photons. The temporal resolution of the all-optical switch is determined by two effects: the pulse duration of the pump and the group velocity difference between the pump and signal, caused by dispersion in the fiber. This group velocity walk-off, which broadens the interaction time here to approximately 2\,ps, is beneficial as it enables uniform switching across the entire signal photon wavepacket.

\begin{figure*}[t!]
	\centering
		\includegraphics[width=0.85\textwidth]{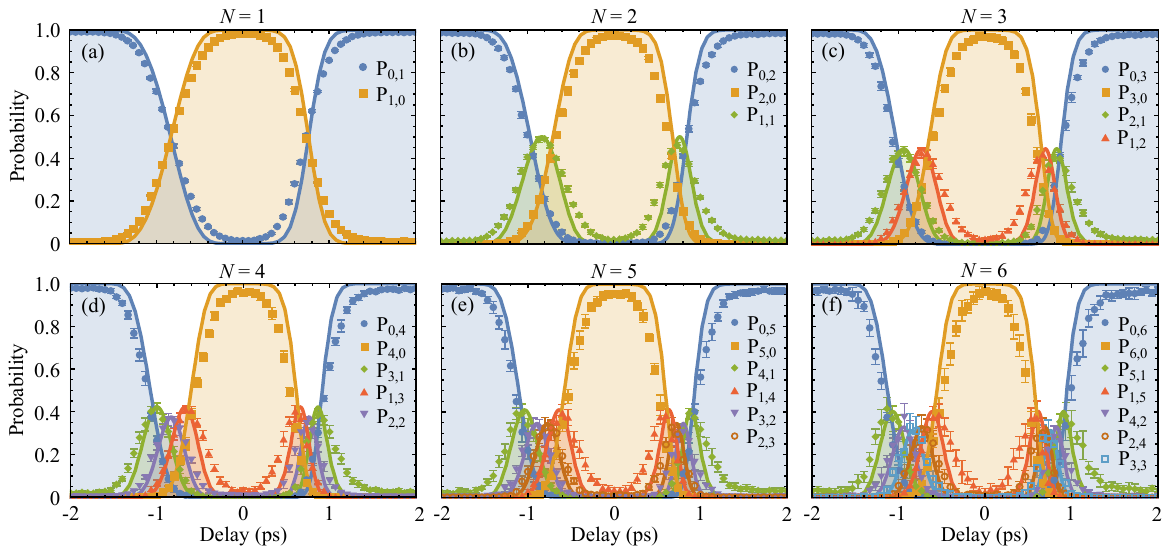}
	\caption{\textbf{Switching of heralded photon-number states.} {The measured probability, $P_{n_S,n_U}$, of detecting $n_S$ photons at $D_S$ and $n_U$ photons at $D_U$, heralded by measurement of $N$ photons at $D_i$, where $n_S+n_U=N$, is shown in (a)--(f) for $N=$ 1--6. {Probabilities are shown as a function of pump pulse delay, where pump pulse energy is held constant at 8\,nJ.} Experimental data (denoted by markers with error bars to one standard deviation) are compared to simulation results (lines with area shaded below). Note that in some cases, error bars are smaller than the data points. TESs are used for all measurements presented here.}}
	\label{fig:TESdata}
\end{figure*}

After the 24\,cm SMF, the 1030\,nm pump pulses are spectrally filtered and a PBS splits the switched and unswitched light into two detection channels, $D_S$ and $D_U$, respectively. {Further details on photon detection are provided in the End Matter.} {The measured insertion loss of the switch, which includes all relevant optical components to achieve near-unity switching (see Supplemental Material), is $2.27\pm0.08$\,dB.} {This loss is primarily limited by fiber coupling, which is not a consideration in the quoted low insertion loss of previous all-fiber based implementations~\cite{purakayastha2022ultrafast, hall2011all}. Except for switching in the short SMF, our entire setup is in free-space. This minimizes dispersion to achieve the fastest possible switching time. Nevertheless, we anticipate that this loss could be further reduced, \textit{e.g.,} with lower loss optical elements and automated coupling to the SMF, or perhaps by operating the switch in bulk crystal~\cite{kupchak2017time}.} 

In order to assess the performance of our switch, we introduce the switching efficiency, $\eta$, given by, 
\begin{eqnarray}
    \eta=\sin^2 (2\theta) \sin^2 \left( \frac{\Delta \phi}{2} \right),
\end{eqnarray}
where $\theta$ is the angle between the polarization of the signal and the pump, {$\Delta \phi = 8 \pi n_2 L_\mathrm{eff} I_\mathrm{p} / 3 \lambda_\mathrm{signal} $} is the nonlinear phase shift induced by the pump in the SMF, $n_2$ is the nonlinear refractive index of the SMF, $L_\mathrm{eff}$ is the effective length of the nonlinear medium, $I_\mathrm{p}$ is the intensity of the pump pulse, and $\lambda_{\mathrm{signal}}$ is the signal wavelength. Switching efficiency is maximized when $\theta=\pi/4$ and $\Delta \phi = \pi$. Although this analytical expression provides a simple means of predicting switching efficiency for a given set of input parameters, it does not account for the more complicated interplay between self-phase modulation (SPM) and group velocity dispersion (GVD) effects on the pump pulse as it propagates through the nonlinear switching medium. {We introduce a more comprehensive simulation in this work, based on a split-step Fourier method approach~\cite{agrawal2013nonlinear} and described in detail in the Supplemental Material.}

The experimental switching efficiency is determined by measuring the coincidence rates between one idler photon and its corresponding switched or unswitched signal photon as
\begin{eqnarray}
    \eta_\mathrm{exp}=N_{S,i}/(N_{U,i}+N_{S,i}),
\end{eqnarray}
where $N_{S,i}$ and $N_{U,i}$ represent the coincidence rates between the idler photon and the signal photon exiting the switched and unswitched output ports of the PBS, respectively. {This switching efficiency is equivalent to a measure of extinction ratio and is independent of insertion loss.} The switching efficiency, measured with the SNSPDs for a mean photon number of $0.24\pm0.02$, is shown in Fig.~\ref{fig:SNSPDdata}(a), as a function of pump pulse energy and relative delay between the pump and signal, $\Delta\tau$. 

Experimental measurements are compared to our comprehensive simulation, as seen in Fig.~\ref{fig:SNSPDdata}(b), which uses the split-step Fourier method to solve the generalized pulse propagation equation. We model the evolution of both the strong pump and weak signal photons, based on the dispersion curve provided by the vendor for the 24\,cm 1060XP SMF. Further details on the simulation are provided in the Supplemental Material. We first validate our model by comparing the simulated output spectra with the experimentally measured spectra, confirming the expected self-phase-modulation-induced spectral broadening of the pump pulses as a function of their energy. With this validated model, the detailed propagation of the pump and signal photon wavepackets through the SMF can be simulated, along with the deconvoluted switching response. These simulations provide insight into the dynamics driving the observed switching behavior and allow us to analyze the nonlinear and dispersive effects that influence the switching response. 

For a fixed delay of $\Delta\tau=0$\,ps, as seen in Fig.~\ref{fig:SNSPDdata}(c), the switching efficiency increases with pump pulse energy, following the expected quadratic sinusoidal curve. For a pump pulse energy of 8\,nJ, we achieve a switching efficiency of $>$99\%. Beyond this point, the efficiency decreases, as the nonlinear phase shift induced by the pump exceeds the point at which $\Delta \phi = \pi$. This behaviour is observed in both the experimental data and simulation results, which show good qualitative agreement. {Notably, we observe very low noise: approximately $1\times10^{-5}$ noise counts per pump pulse in each of the signal detection ports, which is an order of magnitude smaller than in previous demonstrations in the visible regime~\cite{kupchak2019terahertz}. This noise can be attributed to other undesired nonlinear processes induced by the pump within the switch.} Given that the overall probability of heralded detection of the signal photon is 32\% (which includes both the heralding efficiency and full system efficiency), we can infer a signal-to-noise ratio of 32,000.

An important characteristic of our all-optical switch is its temporal resolution, crucial for enabling ultrafast time-bin encoding~\cite{bouchard2022quantum, bouchard2023measuring, bouchard2024programmable, fenwick2024photonic} and other ultrafast quantum photonic applications in the telecom band. The temporal profile of the switch can be seen in Fig.~\ref{fig:SNSPDdata}(d), where switching efficiency is plotted as a function of delay $\Delta\tau$, for a fixed pump pulse energy of 8\,nJ. Again, we see good qualitative agreement between the experimental measurements and simulation results. {The switching resolution, determined by considering the full width at 10\,dB, is found to be 2.3\,ps, which corresponds to a bandwidth of approximately 434\,GHz.} {Considering the relatively low insertion loss, this compares favourably to state-of-the-art electro-optic modulators which have seen bandwidths $>100$\,GHz~\cite{weigel2018bonded}.} {Notably, the top of the curve remains flat near 100\% switching efficiency ($>98$\% over a span of about 533\,fs), indicating that the switch is insensitive to small variations in pump arrival time within this region.} This flatness suggests that the pump pulse has fully traversed the signal photon wavepacket in the SMF, imprinting a nearly uniform temporal phase profile. Such uniformity is essential for achieving near-unity switching efficiency.

Given the high efficiency performance of the switch in the single-photon regime, as measured with the SNSPDs, we next investigate its ability to operate on heralded photon-number states, as measured with TESs. As shown in Figs.~\ref{fig:TESdata}(a)--(f), we measure $N=$ 1--6 idler photon at $D_i$, to herald the detection of $N=$ 1--6 signal photons at the output of the switch, monitoring the probability of each possible outcome, $P_{n_S,n_U}$, where $n_S$ and $n_U$ are the number of photons measured at $D_S$ and $D_U$, respectively. {Further details on these results are provided in the End Matter.}

The results presented in this work showcase the viability of using the optical Kerr effect in SMF for all-optical switching of photon-number states at picosecond timescales in the telecom band. The demonstrated near-unity switching efficiency, {combined with reasonably low insertion loss} and low noise levels, underscores the practicality of this approach for ultrafast quantum photonic applications, such as time-bin encoding. Further, the ability to switch heralded-photon number states at these ultrafast timescales opens the door to next-generation quantum technologies which are anticipated to rely on photon number as a resource. The interplay between group velocity mismatch and self-phase modulation in the fiber plays a critical role in shaping the temporal profile of the switch, as supported by both experimental data and numerical simulations. Investigating alternative fiber geometries or nonlinear media may also lead to greater control over the spectral and temporal characteristics of the switch. This platform represents a promising step toward scalable quantum photonic systems, where precise control over photon-number states on ultrafast timescales will enable advances in quantum communication, computation, and metrology.

\section*{Acknowledgments}
We thank Khabat Heshami, Aaron Goldberg, Yingwen Zhang, Ramy Tannous, Nicolas Couture, Noah Lupu-Gladstein, Jonathan Baker, Nicolas Dalbec-Constant, Nathan Roberts, Milica Banic, Denis Guay, Doug Moffatt, and Rune Lausten for support and insightful discussions.

\providecommand{\noopsort}[1]{}
%


\section*{End matter}

\subsection*{Detection}

The signal and idler photons are detected with either superconducting nanowire single-photon detectors (SNSPDs) or transition-edge sensors (TESs), where PNR capabilities are accessible in both configurations. The SNSPDs used here have a low jitter time of $\sim$20\,ps, which can be leveraged to separate one-photon detection events from multi-photon detection events. By operating the SNSPDs with an appropriate trigger level, one-photon events will trigger a detection later than two-photon events, and so on. This can be seen in Fig.~\ref{fig:experimental}(b), where the bright spot, circled by the dashed white outline, corresponds to coincidence events in which one photon is measured in both the signal and idler channels. When operating the switch with SNSPDs, we consider only coincidences measured in the 60\,ps window enclosed by the dashed white outline to filter out any detection events arising from more than one heralded photon. It is important to note that the PNR capabilities of the SNSPDs are only accessible when operating on ultrafast timescales, as demonstrated here. 

On the other hand, the TESs provide photon counting capabilities up to higher photon numbers. This allows us to operate our switch on specific photon-number states, which is a useful prospective in the context of photonic quantum information processing. The joint photon-number distribution of the source, as measured with the TESs, is presented in Fig.~\ref{fig:experimental}(c) for characterization purposes. Note that switching demonstrations with the SNSPDs and TESs were performed in the low ($\braket{n}=0.24\pm0.02$) and high ($\braket{n}=3.86\pm0.05$) mean photon number regimes, respectively. The joint photon-number intensity distribution for a range of mean photon numbers is included in the Supplemental Material. Importantly, the spectrum of switched photons is not changed by the switching process, as demonstrated in Fig.~\ref{fig:experimental}(d). {Given that the switch relies on XPM, which is independent of the input state intensity, this is anticipated to hold true in the high-gain regime of parametric downconversion as well.} Here, we achieve spectrally resolved detection of the signal photons with a dispersion compensating module (SMFDK-S-060-03-10 from OFS), which has a group delay dispersion of $D=1033$\,ps/nm, and the SNSPDs to implement a time-of-flight spectrometer~\cite{avenhaus2009fiber}. The spectrally resolved signal photons are heralded by detection of the idler photon.

\subsection*{Heralded photon-number state switching}

{It is important to note that even if initial state exiting the SPDC crystal has perfect photon-number correlations between the signal and idler, the system loss will degrade this correlation. As such, the results presented here do not imply the achievement of near-deterministic Fock state preparation.} {When interpreting the data presented in Fig.~\ref{fig:TESdata}, the switch can be taken as a beamsplitter with a variable reflectivity (equivalent to the switching efficiency) that depends on the pump pulse energy and delay, where an $N$-photon Fock state enters at one input port and a vacuum state enters at the other input port. Here, we opt to hold pump pulse energy constant at 8\,nJ and scan pump pulse delay. The output probabilities then take the form
\begin{eqnarray}
\label{eq:binom}
    P_{n_S,n_U}= \binom{N}{n_S} \eta^{n_S} (1-\eta)^{N-n_S},
\end{eqnarray}
where $\eta=\eta(\Delta\tau)$ defines the splitting ratio as a function of pump pulse delay, and for a fixed $N$, $\sum_{i=0}^N P_{n_S=i,n_U=N-i} = 1 $. Experimental measurements of $P_{n_S,n_U}$ are compared with simulated values based on the simulated switch efficiency in Fig.~\ref{fig:SNSPDdata}(d) and Eq.~\ref{eq:binom}, where we see good qualitative agreement.} Furthermore, we observe that the $P_{N,0}$ curve becomes narrower for larger $N$, which is indicative of the anticipated increasing sensitivity to the switch splitting ratio.

Importantly, we still observe a high switching efficiency for $\Delta\tau=0$ for large $N$, as summarized in Table~\ref{tab:PN0}. {The slightly lower switching efficiencies observed with the TESs can be explained by the longer detection time window, \textit{i.e.,} 1\,$\mu$s for the TESs instead of $60$\,ps for the SNSPDs, which results in additional noise in the data.} 

\setlength{\tabcolsep}{36pt}
\setlength{\extrarowheight}{0.02pt}
\begin{table}[h!]
  \centering
\begin{tabular}{c r} 
 \hline\hline
 \rule[-2ex]{0pt}{5ex}$N$ photons & $P_{N,0}(\Delta\tau=0)$ \\
 \hline\hline

 \rule[-2ex]{0pt}{5ex}1 & $0.985\pm0.001$ \\
 \rule[-2ex]{0pt}{5ex}2 & $0.977\pm0.003$ \\
 \rule[-2ex]{0pt}{5ex}3 & $0.970\pm0.006$ \\
 \rule[-2ex]{0pt}{5ex}4 & $0.958\pm0.009$ \\
 \rule[-2ex]{0pt}{5ex}5 & $0.955\pm0.013$ \\
 \rule[-2ex]{0pt}{5ex}6 & $0.965\pm0.013$ \\

 \hline

\end{tabular}
  \caption{\textbf{Switching efficiency of heralded \textit{N}-photon state.} Efficiencies for $N=$ 1--6 heralded photon-number states for a pump delay of $\Delta\tau=0$.}
  \label{tab:PN0}
\end{table}

Nevertheless, these results demonstrate the ability of our switch to reliably act on heralded photon-number states, opening the door to continuous-variable encoding schemes in, \textit{e.g.,} quantum information processing with ultrafast time bins. Furthermore, the switching of Fock states is anticipated to allow for the injecting of nongaussianity into photonic quantum circuits~\cite{crescimanna2024seeding}, a useful capability in the context of quantum information processing. Heralded photon-number states are the next stage in the evolution of photonic quantum states, providing expanded capabilities in comparison to their predecessors (\textit{i.e.,} attenuated coherent states or two-mode squeezed vacuum without PNR)~\cite{afek2010high,konno2024logical,takase2021generation,endo2025high}.

\end{document}